\documentclass[draft]{aipprocMHTE}
\layoutstyle{6x9}

\newcommand{\ket}[1]{|#1 \rangle}

\newcommand{\pf}{\rm{^{40}K}}  
\newcommand{\re}{\rm{^{87}Rb}}

\usepackage{graphicx}	
\usepackage{amssymb}	

\begin{document}

\title{Dual-species quantum degeneracy of $\mathbf{\pf}$ and
$\mathbf{\re}$ on an atom chip}


\author{M.~H.~T.~Extavour, L.~J.~LeBlanc, T.~Schumm, B.~Cieslak, 
S.~Myrskog, A.~Stummer, S.~Aubin, J.~H.~Thywissen}
{ address={Department of Physics and Institute for Optical
Sciences, University of Toronto, 60 St.
George Street, Toronto, Canada} }
%

\begin{abstract}
In this article we review our recent experiments with a
$\pf$-$\re$ mixture. We demonstrate rapid sympathetic cooling of
a $\pf$-$\re$ mixture to dual quantum degeneracy on an atom chip.
We also provide details on efficient BEC production,
species-selective magnetic confinement, and progress toward
integration of an optical lattice with an atom chip. The
efficiency of our evaporation allows us to reach dual degeneracy
after just 6 s of evaporation - more rapidly than in conventional
magnetic traps.  When optimizing evaporative cooling for
efficient evaporation of $\re$ alone we achieve BEC after just 4
s of evaporation and an 8 s total cycle time. 
\end{abstract}

\maketitle


\section{Introduction}
Ultra-cold gases of neutral fermionic atoms are of great interest
in atomic and condensed matter physics, yet they remain
challenging to produce experimentally.  This article describes
the production and manipulation of ultra-cold fermionic and
bosonic gases using an atom chip and a simple, single-chamber
vacuum system. We describe the observation of efficient
evaporation to a Bose-Einstein condensate (BEC) in $\re$, along
with the rapid sympathetic evaporative cooling of a $\pf$-$\re$
mixture to dual degeneracy - a BEC of $\re$ and a degenerate
Fermi gas (DFG) of $\pf$.  We discuss the use of radio frequency
(RF) dressed potentials to create species-selective magnetic
microtraps.  Finally, we detail progress toward incorporating an
optical lattice into our experiment.  Although some of the
details of our setup and DFG-BEC have been reported
elsewhere~\cite{Aubin:JLTP,Aubin:DFG}, this article provides an
overview and update on important figures of merit.

The article begins with an outline of our initial optical and
magnetic trapping and cooling stages for $\re$ and $\pf$.  Next,
we give a description of our observation of quantum degeneracy in
both species, and further details of our evaporative cooling
trajectories.  We then describe the use of RF-dressed adiabatic
potentials to produce a species-selective microtrap.  The article
ends with a discussion of our plans to integrate optical lattices
into the experiment, and a discussion of the technical
improvements being made to our atom chip.

%
\section{LOADING AND TRAPPING}
We begin by trapping neutral atoms directly from
atomic vapour with a six-beam magneto-optical trap (MOT). The
dual species MOT consists of two overlapping sets beams: six
counterpropagating beams at 780 nm to trap $\re$, and six beams
at 767 nm for $\pf$ \cite{Aubin:JLTP}. With $\re$ MOT loading
times as short at 950 ms, we can achieve BEC . Working with the
$\pf$-$\re$ mixture requires longer loading times due to the low
isotopic abundance of $\pf$ in our dispensers\footnote{We use KCl
powder whose $\pf$ isotopic abundance has been enriched to 5\%
from the naturally-occurring 0.012\%.  For comparison, the
natural abundance of $\re$ is 28\%.}:
we first load $\pf$ alone into the MOT for 10-20 s, after which
$\re$ is loaded for an additional 1-5 s, while maintaining the
$\pf$ population.  Both sets of MOT beams operate with a detuning
of -26 MHz until the last 10 ms of the load, when the $\pf$ is
compressed by reducing the detuning of the $\pf$ beams to -10
MHz.  With this procedure we load 10$^9$ $\re$ atoms and 10$^7$
$\pf$ atoms into the MOT.  

\begin{figure}
\includegraphics[height=4cm]{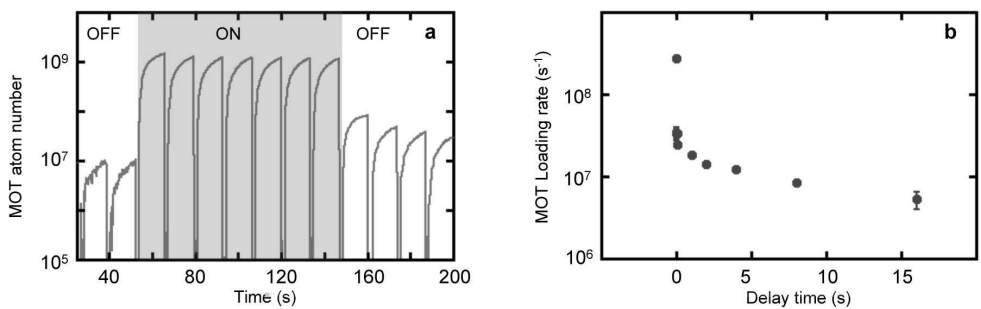} 
\caption{{\bf MOT loading via LIAD} 
We load the MOT from the atomic vapour formed via light-induced
atomic desorption (LIAD) from the interior walls of our Pyrex
vacuum cell. (\textbf{a}) As the $\re$ MOT is repeatedly loaded
from atomic vapour, we observe a one-hundred-fold increase in
atom number (greyed area), which is lost after the LEDs are
switched off. (\textbf{b}) The MOT loading rate as function of
time \emph{after} the LEDs have been switched off.  Once the LEDs are
off, the MOT loading rate undergoes a sharp initial decrease from
the LED-on value of 3 $\times$ 10$^8$ s$^{-1}$, followed by a
slower decay to the LED-off value of \mbox{$\sim$ 10$^6$
s$^{-1}$} in approximately 300 s.} 
\label{fig:LIAD} 
\end{figure}

The background atomic $\re$ and $\pf$ vapours are enhanced via
light-induced atomic desorption (LIAD) of alkali atoms from the
inner walls of a 75 mm $\times$ 75 mm $\times$ 165 mm rectangular
Pyrex vacuum cell \cite{Aubin:JLTP}.  By illuminating the cell
using approximately \mbox{600 mW} of incoherent \mbox{405 nm} LED
light\footnote{We use ten Epitex L405 series UV illuminators.},
we observe a 100-fold increase in $\re$ MOT atom number compared
to loading from the background vapour alone (see Figure
\ref{fig:LIAD}).  The increase in $\pf$ MOT atom number is
unknown; we cannot observe a $\pf$ MOT without LIAD. A coating of
$\re$ and $\pf$ atoms on the interior walls of our vacuum chamber
is replenished every few weeks by running a commercial Rb
dispenser and a home-made K dispenser.

After MOT loading, 3 ms of optical molasses cooling is applied to
the $\re$.  Finally, both species are optically pumped into
weak-field-seeking hyperfine ground states: $\ket{F=2,m_F=2}$ for
$\re$ and $\ket{F=9/2,m_F=9/2}$ for $\pf$.

We use a magnetic transfer procedure to transport the
atoms from the site of the MOT to the atom chip, the site of
evaporative cooling. Immediately following optical pumping, all
optical fields are extinguished and atoms are confined in a
two-coil quadrupole magnetic trap formed at the site of the MOT.
A biasing magnetic field generated by a third coil shifts the
magnetic trap centre vertically to just below the surface of the
atom chip, which is mounted horizontally in the vacuum chamber 5
cm above the MOT site.  

Once near the surface of the atom chip, atoms are smoothly
transferred from the quadrupole magnetic trap into an anisotropic,
Ioffe-Pritchard-type magnetic microtrap. The chip trap is formed
by combining the static magnetic fields generated by DC current
in a `Z'-shaped wire on the atom chip with an external, uniform
magnetic field parallel to the chip surface \cite{Reichel:chips}.
This combination produces a non-zero magnetic field minimum just
below the chip surface, in which atoms in weak-field-seeking
hyperfine states are confined.  Atoms are transferred into the
microtrap by ramping up the Z-wire current and external field
while ramping down the quadrupole trap.  
%

\section{QUANTUM DEGENERACY}
We cool to quantum degeneracy using forced RF evaporative cooling
of $\re$.   Whether cooling $\re$ alone or a $\pf$-$\re$ mixture
\cite{Roati:mixture,Jin:BECDFG,Sengstock:BECDFG}, RF radiation
acts only on the $\re$, ejecting hot atoms from the magnetic trap
and allowing the remaining atoms to rethermalize to a new, lower
temperature.  The the process of indirectly cooling the one
species by direct RF evaporation of the other is referred to as
sympathetic cooling
\cite{WK:sympathetic,Wieman:symp,Schreck:symp,Inouye:fesh,Ferlaino:KRb}.
In our case $\pf$ is cooled indirectly via thermalizing
elastic collisions with the $\re$.

\begin{figure}
$\begin{array}{c@{\hspace{5mm}}c}
  \includegraphics[height=4cm]{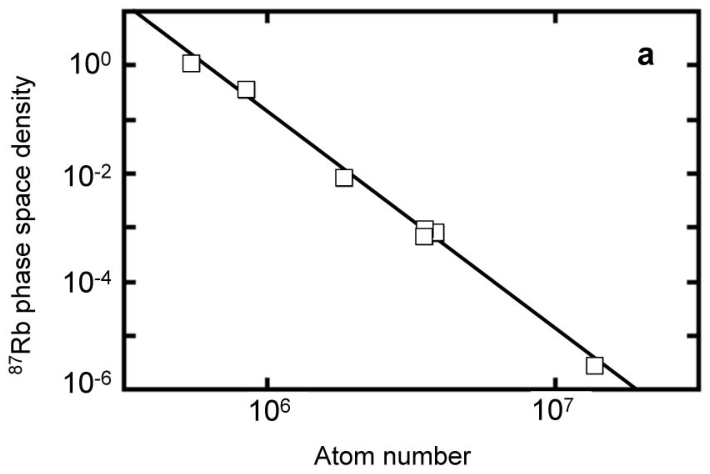} &
  \includegraphics[height=4cm]{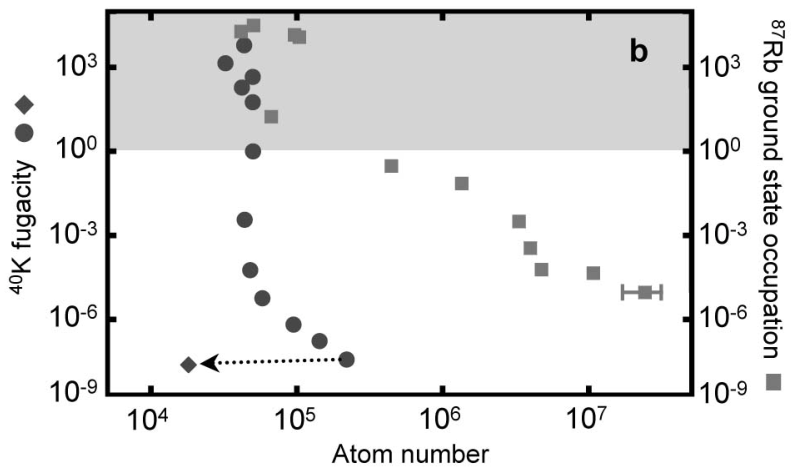}
\end{array}$
  \caption{{\bf Evaporative cooling in a chip trap.} 
(\textbf{a}) The $\re$ evaporative cooling
trajectory optimized for $\re$ alone is more efficient than that
used for sympathetic cooling.  The line of best fit establishes
the log-slope efficiency (defined in the text).  (\textbf{b})
Spin-polarized fermions without a bosonic bath cannot be
evaporatively cooled (diamond). However, if bosonic $\re$
(squares) is evaporatively cooled, then fermionic $\pf$ is
sympathetically cooled (circles) to quantum degeneracy (greyed
area). For bosonic $\re$, the vertical axis is the occupation of
the ground state; for fermionic $\pf$, the vertical axis is the
fugacity. These two quantities are equivalent in the
non-degenerate limit. A typical run-to-run spread in atom number
is shown on the right-most point; all vertical error bars are
smaller than the marker size.}  
\label{fig:PSD} 
\end{figure}

\paragraph{Evaporative cooling of bosons alone}
In the $\re$-only case we achieve a BEC of \mbox{2 $\times$
10$^5$} atoms when empirically optimizing for evaporation
efficiency.  We first load the MOT for 2 s, magnetically trap and
transfer to the chip in 950 ms, and load atoms into a chip
microtrap in 310 ms at $T \sim$ 300 $\mu$K. Next, we perform
forced RF evaporative cooling in which the RF frequency
$\nu_{RF}$ is swept from 15.0 MHz to 3.6 MHz in 4.3 s, bringing
the the total cycle time to $\sim$ 7.6 s.  Evaporation is carried
out in an anisotropic trap with trap frequencies of $\omega_\perp
/ 2\pi$ = 560 Hz and $\omega_\ell / 2\pi$ = 32 Hz. The log-slope
efficiency $\eta_{evap}$ of the evaporation, defined as $-
\partial [log(\rho)]/\partial [log(N)]$, where $\rho$ is the
phase space density\footnote{Phase space density is defined as
$n_0 \lambda_{dB}^3$ in this context, where $n_0$ is the peak
atomic density, and $\lambda_{dB}$ the thermal deBroglie
wavelength.} and $N$ the atom number, is equal to 4.0 $\pm$ 0.1
when we optimize evaporation for $\re$ alone.  For comparison,
\mbox{$\eta_{evap}$ = 2.9 $\pm$ 0.4} when evaporation is
optimized for sympathetic cooling to dual degeneracy (see Figure
\ref{fig:PSD}). Our largest BECs are achieved with cycle times of
15 s, and contain \mbox{3 $\times$ 10$^5$} atoms.

When optimizing for minimum cycle time, we sacrifice some
evaporation efficiency, but achieve a BEC of 2 $\times$ 10$^4$
atoms with cycle times as short as 5.7 s. In this case, we load
the MOT for 950 ms, and transfer to the atom chip as described
above.  Evaporative cooling is carried out in two empirically
optimized steps. The initial 700 ms RF frequency sweep from
$\nu_{RF}$ = 33.5 MHz to $\nu_{RF}$ = 6.0 MHz is carried out in
a compressed trap, which has $\omega_\perp / 2\pi$ = 2.8 kHz at
its centre and a transverse magnetic field gradient of 4800 G/cm
at its edge.  Next, the microtrap is decompressed until
$\omega_\perp/2\pi = 861 \,$ Hz and $\omega_\ell/2\pi=32$\,Hz.  A
second evaporative $\nu_{RF}$ sweep is carried out here, from
27.9 MHz to 3.5 MHz in 2030 ms.  

%

\begin{figure}[h]
  \includegraphics[height=5cm]{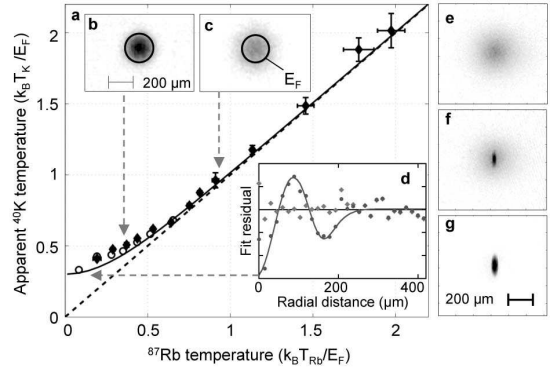}
  \caption{ {\bf Observation of Fermi statistics.} Due to Fermi
pressure, Fermi degenerate $\pf$ clouds appear to stop getting
colder, even when the reservoir temperature approaches zero.
(\textbf{a}) The apparent temperature of the fermions, as measured
by Gaussian fits to images of $\pf$ clouds, is plotted versus
temperature of both thermal (diamonds) and Bose-condensed
(circles) $\re$. Data is compared to a Gaussian fit of
a theoretically generated ideal Fermi distribution (solid line) and
its classical expectation (dashed line).  Both temperatures are
scaled by the Fermi energy $E_\mathrm{F}$ of each $\pf$ cloud.
Error bars are statistical (one standard deviation), with
uncertainty smaller than the sizes of symbols for lower
temperature data.
Absorption images are shown for $k_\mathrm{B}
T/E_\mathrm{F}=0.35$ (\textbf{b}) and $0.95$ (\textbf{c}),
including a circle indicating the Fermi energy $E_\mathrm{F}$.
(\textbf{d}) A closer look at the fermion cloud shape reveals that
it does not follow a Boltzmann distribution. The fit residuals of
a radially averaged cloud profile show a strong systematic
deviation when assuming Boltzmann (circles) instead of Fermi
(diamonds) statistics. A degenerate Fermi cloud is flatter at its
centre than a Boltzmann distribution, and falls more sharply to
zero near its edge.
(\textbf{e-g}) Absorption images of $\re$ atoms after 20 ms of
time-of-flight expansion: above (\textbf{e}), just below
(\textbf{f}), and well below $T_c$ (\textbf{g}), which shows 2.5
$\times$ 10$^5$ atoms in a quasi-pure BEC.  $T_c \approx  400$
nK in this data.  The BEC is anisotropic in-trap, with its long
axis aligned horizontally.  This alignment is inverted at long
times of flight as evident in these images.}
\label{fig:signatures} 
\end{figure}

We detect the phase transition from thermal gas to BEC by
tracking the peak phase space density  of the gas.
We fit time-of-flight absorption images with a composite peak
function to capture the separate behaviour of the condensed and
non-condensed fractions: a Thomas-Fermi parabola describes the
condensate fraction, and a Bose function non-condensed
fraction (see Figure \ref{fig:signatures}f).  These fits yield
reliable atom number and temperature data, from which we
determine the phase (thermal atoms or BEC) of the cloud.  Once
below $T_c$ we also observe an inversion in the aspect ratio of
the expanding cloud \cite{WK:MPU}.  Figures
\ref{fig:signatures}e-g show absorption images of expanding
condensates at three successively colder temperatures: above
$T_c$, where the gas is entirely non-condensed; near $T_c$, where
the BEC begins to emerge; and much below $T_c$, where the
condensed fraction dominates the cloud. 

%
%
%
\paragraph{Sympathetic cooling of the Bose-Fermi mixture} When
working with the $\pf$-$\re$ mixture, we find that RF sweep
times faster than $\sim$ 6 s are not successful in achieving dual
degeneracy.  This is because the $\pf$-$\re$ rethermalization
time lags that of $\re$-$\re$ at high temperatures. This
empirical observation was studied in the context of the
Ramsauer-Townsend effect in \cite{Aubin:DFG}.  We are also unable
to sympathetically cool in the compressed trap used for rapid BEC
production, since the higher densities in this trap lead to
three-body loss, preventing efficient cooling of the $\pf$-$\re$
mixture~\cite{Roati:mixture,Sengstock:BECDFG}.  Figure
\ref{fig:PSD} shows the relative evaporation efficiencies $\re$
and $\pf$ during sympathetic cooling, as well as during
single-species evaporation of $\re$ alone. 

%

We first load $2\times10^5$ $\pf$ and $2\times10^7$ $\re$ doubly
spin-polarized atoms into a 1.1 mK-deep microtrap at a
temperature $\gtrsim 300$\,$\mu$K. The $\pf$ radial and
longitudinal trapping frequencies are measured to be
$\omega_\perp/2\pi = 826\pm7$\,Hz and
$\omega_\ell/2\pi=46.2\pm0.7$\,Hz, respectively\footnote{The
corresponding $\re$ trap frequencies are a factor of
$\sqrt{m_{Rb}/m_K} \approx 1.47$ smaller than those for $\pf$,
where $m_\mathrm{Rb}$ and $m_\mathrm{K}$ are the atomic masses of
$\re$ and $\pf$, respectively.}. $\nu_{RF}$ is swept from 28.6
MHZ to 3.6 MHz in 6.15 s. 

As with $\re$, we assess the degree of quantum degeneracy in
$\pf$ by fitting absorption data with ideal gas theory. In the
case of $\pf$, fits using ideal fermion density distributions yield
temperature, density, Fermi energy and fugacity. However,
since there is no phase transition to the quantum degenerate
state in cold, spin-polarized non-interacting fermions,
distinguishing degenerate from non-degenerate Fermi
gases is much more challenging.  Nevertheless, we observe a clear
signature of Fermi degeneracy in $\pf$ below $T\approx1$\,$\mu$K
in the shape of the time-of-flight distribution.  

We compare the rms cloud sizes of the $\pf$ and the non-condensed
$\re$ fraction by fitting the density profiles to the Gaussian
shapes predicted by Boltzmann statistics.
Figure~\ref{fig:signatures}a shows that the apparent
(i.e., Gaussian-estimated) $\pf$ temperature approaches a finite
value while the $\re$ temperature approaches zero, even though
the two gases are in good thermal contact.  This deviation is
evidence of the Fermi pressure expected of a gas obeying
Fermi-Dirac statistics, and of the Pauli exclusion
principle~\cite{Truscott:FP}: at zero temperature, fermions fill
all available energy states up to the Fermi energy $E_\mathrm{F}
= \hbar (6 N \omega_\perp^2 \omega_\ell)^{1/3}$, where $N$ is the
number of fermions.  We plot data with thermal (diamonds) and
Bose-condensed (circles) $\re$ separately, to show that the
density-dependent attractive interaction between $\pf$ and $\re$
does not significantly affect the release energy.  For our
typical parameters,  $E_\mathrm{F} \approx k_\mathrm{B} \times $
1.1\,$\mu$K.  After all $\re$ atoms have been evaporated, we use
Fermi-Dirac fits to measure temperature, and find $k_\mathrm{B}
T/E_\mathrm{F}$  as low as $0.09\pm 0.05$ with as many as
$4\times 10^4$ $\pf$ atoms.


%
\section{Species-selective microtraps}
A common practice among experimental realizations of $\pf$-$\re$
mixtures \cite{Roati:mixture,Sengstock:BECDFG,Jin:BECDFG} (and
Bose-Fermi mixtures generally) is the use of a single potential
which confines both species.  The ability to separately
control the external potential of each species is a new and
useful tool for studying such mixtures, which we demonstrate
here.  Tunable, species-selective magnetic confinement of our
$\pf$-$\re$ mixture is accomplished by combining a transverse
field oscillating at radio frequency (RF) with the static
magnetic microtrap~\cite{Zobay:RF,
Colombe:RF,Schumm:doubleBEC,Courteille:RF,DeMarco:RF}.

When the RF frequency is resonant with the magnetic hyperfine
splitting of one atomic species, the RF radiation induces a
coupling between the $m_F$ levels in that species, just as in
forced RF evaporative cooling.  Unlike in evaporative cooling,
however, we ramp the RF frequency from \emph{below} to
\emph{above} resonance.  Doing so dynamically transforms the
potential from the $\ket{F=2,m_F=2}$ single-well ($\nu_{RF}$
below resonance) into a dressed state double-well ($\nu_{RF}$
above resonance)\footnote{In evaporative cooling, the RF
frequency is ramped down starting from \emph{above} resonance, so
that atoms experience a single-well dressed state potential at
all times during evaporation.}. The central barrier height and
well separation of the double-well are controlled by the RF
frequency and amplitude~\cite{Schumm:doubleBEC}.

\begin{figure}
  \includegraphics[height=3cm]{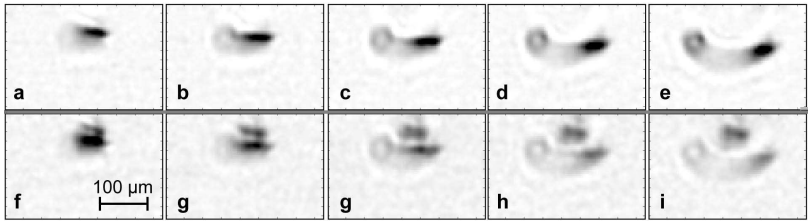}
  \caption{
{\bf Species-selective manipulation.} A static chip trap loaded
with a $\pf$-$\re$ mixture is coupled to an RF field
to induce a double-well dressed potential for $\re$ only;  the
$\pf$ portion of the mixture remains unsplit.  The final RF
coupling frequency increases from 4.0 MHz (\textbf{a} and
\textbf{e}) to 6.0 MHz (\textbf{f} and \textbf{i}) in steps of
0.5 MHz, deforming the $\re$ dressed potential minimum.
(\textbf{a-e}) Absorption images of the mixture with probe light
resonant to $\re$ only.  (\textbf{f-i}) Absorption images of the
mixture taken with two-colour probe light resonant to both
species. Both species are visible, with the $\pf$ atoms nearly
unaffected by the RF dressing.}
\label{fig:specSelect} \end{figure}

We achieve species-selective dressed potentials when the RF
radiation is resonant to the ground state hyperfine splitting of
$\re$, but well detuned from that of $\pf$. In a magnetic field
the energy separation between the five $F=2$ hyperfine levels in
the $5^2S_{1/2}$ $\re$ ground state is roughly twice the
separation of that between the ten $F=9/2$ hyperfine levels in
the $4^2S_{1/2}$ $\pf$ ground state\footnote{$U_{magnetic} = m_F
g_F \mu_B B$, where $\mu_B$ is the Bohr magneton and $B$ the
magnitude of magnetic field.  The Land{\'e} g-factor $g_F$ is
$2/9$ for $F$=9/2 in $\pf$ and $1/2$ for $F=2$ in $\re$.}.  The
$\re$ double well is formed by increasing the RF frequency from
3.6 MHz to a final RF frequency between 4.0 MHz and 6.0 MHz (see
Figure \ref{fig:specSelect}).  The single-well $\re$ static trap
deforms into a double-well as the RF frequency passes through the
$\re$ trap bottom, located at 3.8 MHz, while the $\pf$ potential
remains almost unaffected by the RF dressing: we estimate the
Rabi frequency $\Omega_R$ to be as high as 1 MHz, whereas the
static magnetic bias field provides a minimum detuning of 2.1 MHz
between $\pf$ and $\re$ resonances.
%
%
%

Figure \ref{fig:specSelect} shows absorption images of the
$\pf$-$\re$ mixture 0.3 ms after being released from the
RF-dressed trap. As expected, the $\re$ atoms split primarily
into two clouds as the double-well RF frequency is increased. The
$\pf$ atoms, still experiencing the bare static trap, remain
confined to a single, central cloud. This species separation was
observed via absorption imaging with a two-colour probe beam, the
two optical frequencies corresponding to the imaging transitions
in $\re$ and $\pf$.

\section{Next generation atom chip: on-chip optical lattices}
With an eye to future experiments, we are in the midst of
developing a new atom chip which can support an optical lattice.
The atom chip will be covered by a dielectric mirror, enabling
the formation of a 1D optical lattice near the chip surface with
a single, retroreflected laser beam. Improved chip optical access
will allow the incorporation of additional beams to form 2D and
3D lattices. 

Mirrorizing is carried out following Reichel's replica technique
\cite{Reichel:mirror}.  Once the chip wires have been fabricated
on the substrate, a planarizing epoxy is applied to the atom
chip.  A dielectric mirror on a glass substrate is then carefully
lowered onto the epoxy layer, pressed, and held in place while
the epoxy cures.  Once the epoxy has set and cured, the mirror
substrate is carefully removed, leaving the smooth dielectric
mirror adhered to the atom chip.  

The new atom chip will also contain smoother wires with higher
current capacity, fabricated with photolithography, evaporative
deposition, and lift-off.  The chip substrate is an AlN wafer
with filled vias connecting the front (wire) side to the backside
of the chip.  This allows for electrical contacts to be made on
the back side of the chip, removing them as obstacles and thus
greatly increasing our optical access.  After evaporative
deposition of Cr and Ti adhesion layers, a 2.7 $\mu$m-thick layer
of Ag and a thin capping layer of Au are evaporated directly onto
the substrate to form the chip wires.  Evaporative deposition was
chosen over electroplating to reduce the wire roughness, which in
turn will reduce the rugosity of the resulting magnetic
potential~\cite{Fortagh:frag,Esteve:frag,Groth:chipfab}.  High DC
current capacity in these wires is enabled by the combination of
the high electrical conductivity of Ag and the high thermal
conductivity of AlN.  Recent tests have shown that as much as 7.5
A of current can be passed through a 20 $\mu$m-wide, 2.7
$\mu$m-tall, 4.6 mm-long wire; this corresponds to a current
density of greater than 10$^7$ A/cm$^2$.


\begin{theacknowledgments}
We would like to thank M.~H{\"a}ffner, I.~Leroux and D.~McKay for
their early contributions to this work.  We would also like to thank
the Groupe d'Optique Atomique in Orsay for technical
collaborations.  This work is supported by the NSERC, CFI, the
Province of Ontario, CRC, and Research Corporation. S.~A. and
L.~J.~L. acknowledge support from NSERC.  M.~H.~T.~E. acknowledges
support from OGS.  
\end{theacknowledgments}



\bibliographystyle{aipprocl} 

\bibliography{icap06_proc} 

%

\end{document}